\documentclass[6pt]{article}
\newfont{\bbbold}{msbm10 scaled \magstep1}

\newfont{\goth}{eufm10 scaled \magstep1}

\def\gh{\mbox{\goth h}}

\def\a{\alpha}\def\adt{\dot \alpha}
\def\ah{\hat\alpha}
\def\b{\beta}\def\bdt{\dot \beta}
\def\bh{\hat\beta}
\def\gh{\hat\gamma}
\def\lh{\hat\lambda}
\def\c{\gamma}\def\C{\Gamma}\def\cdt{\dot\gamma}
\def\d{\delta}\def\ddt{\dot\delta}
\def\e{\epsilon}\def\ve{\varepsilon}
\def\f{\phi}
\def\g{\gamma}

\def\i{\iota}

\def\l{\lambda}\def\L{\Lambda}
\def\m{\mu}

\def\p{\pi}

\def\r{\rho}

\def\t{\tau}
\def\th{\theta}

\def\beq{\begin{equation}}\def\eeq{\end{equation}}
\def\beqa{\begin{eqnarray}}\def\eeqa{\end{eqnarray}}
\def\barr{\begin{array}}\def\earr{\end{array}}

\def\o{\omega}\def\O{\Omega}
\def\del{\partial}

\def\xz{\times}

\def\nab{\nabla}

\let\la=\label

\def\nn{\nonumber}
\def\bd{\begin{document}}
\def\ed{\end{document}}
\def\ba{\begin{array}}
\def\ea{\end{array}}
\def\bea{\begin{eqnarray}}
\def\eea{\end{eqnarray}}
\def\ft#1#2{{\textstyle{{\scriptstyle #1}\over {\scriptstyle #2}}}}
\def\fft#1#2{{#1 \over #2}}
\newcommand{\be}{\begin{equation}}
\newcommand{\ee}{\end{equation}}
\newcommand{\eq}[1]{(\ref{#1})}
\newcommand{\w}[1]{\\[0.#1cm]}
\def\eqs#1#2{(\ref{#1}-\ref{#2})}
\def\det{{\rm det\,}}
\def\tr{{\rm tr}}
\newcommand{\Section}[1]{\section{#1} \setcounter{equation}{0}}
\newcommand{\hoch}[1]{$\, ^{#1}$}
\newcommand{\kings}
{\it\small Department of Mathematics, King's College, London, UK}
\newcommand{\iftsp}
{\it\small Instituto de Fisica Teorica, State University of S\~ao Paulo, Brasil}
 \makeatletter
\renewcommand\theequation{\thesection.\arabic{equation}}
\@addtoreset{equation}{section} \makeatother

\newcommand{\auth}
{\large N. Berkovits\hoch{1} and P.S. Howe\hoch{2}}

\begin{document}

\hfill{KCL-TH-08-02}

\hfill{IFT-P.004/2008}

\vspace{20pt}

\begin{center}
{\Large{\bf The cohomology of superspace, pure spinors and invariant integrals}}
 \vspace{30pt}

\auth

\vspace{15pt}

\begin{itemize}
\item [$^1$] \iftsp\item [$^2$] \kings
\end{itemize}

\vspace{60pt}

{\bf Abstract}

\end{center}

The superform construction of supersymmetric invariants, which consists of integrating the top component of a closed superform over spacetime, is reviewed. The cohomological methods necessary for the analysis of closed superforms are discussed and some further theoretical developments presented. The method is applied to higher-order corrections in heterotic string theory up to $\a'^3$. Some partial results on $N=2,d=10$ and $N=1,d=11$ are also given.

\pagebreak \tableofcontents \setcounter{page}{1}


\section{Introduction}


Pure spinors have turned out to be extremely useful in supersymmetric field theory and string theory. They were used in an attempt to solve the auxiliary field problem for $N=1, d=10$ super Yang-Mills (SYM) theory \cite{Nilsson:1985cm}, but they made their first appearance in the modern sense in \cite{Howe:1991mf,Howe:1991bx}
where it was shown how one could derive the equations of motion for $d=10, N=1$ SYM and supergravity, including the gauge sector Chern-Simons term, and for $d=11$ supergravity, from the postulate of pure spinor integrability. This amounts to assuming the existence of a BRST operator formed by contracting the pure spinor with a suitable fermionic derivative. In the case of supergravity one has to use either loop superspace or membrane superspace in order that the two- or three-form potentials in $d=10$ or $d=11$ supergravity can be reinterpreted as one-form gauge fields in the functional superspaces. Moreover, in \cite{Tonin:1991ii,Berkovits:1992aa,Sorokin:2002so}, a version of heterotic string theory was given with $N=2$ worldsheet supersymmetry in which pure spinor variables arise naturally. Somewhat later, the pure spinor variables were reinterpreted as ghosts \cite{Berkovits:2000fe} thereby giving rise to a version of superstring theory which can be quantised in a way which preserves spacetime supersymmetry manifestly. In a series of papers (see \cite{Berkovits:2002zk} for a review and references) it was shown how particles and various strings could be formulated in terms of pure spinors with the latter acting as ghosts for both kappa-symmetry and worldsheet reparametrisations.
An attempt was also made to describe the membrane in a similar way but this is as yet not fully understood \cite{Berkovits:2002uc}, while $d=11$ pure spinors have also been used to describe superparticles and topological M-theory \cite{Grassi:2004xr,Grassione:2004aa}. Supergravity constraints and pure spinors in $d=10$ and $d=11$ have recently been discussed  from the perspective of free differential algebras \cite{Fre':2006es,Fre:2008qw,D'Auria:2008ny,Fre':2008qr}.

Pure spinors also arise naturally in the cohomology of superspace, called spinorial cohomology in \cite{Cederwall:2001dx}. Indeed, the spinorial cohomology groups for purely odd forms (i.e. all odd indices), turn out to be isomorphic to the pure spinor cohomology groups if we adopt the definition that a pure spinor $\l$ satisfies $\l\c_a \l=0$. This coincides with the usual definition in $d=10$ but not in $d=11$ where a pure spinor in Cartan's sense also obeys $\l\c_{ab}\l=0$. There are also spinorial cohomology groups with an additional vector index \cite{Cederwall:2001dx}, relevant to the deformations of the dimension zero torsion, and groups for forms with mixed even and odd indices \cite{Howe:2003cy}. The main application of spinorial cohomology groups has been to the analysis of allowed deformations of theories with maximal supersymmetry for which it is believed that there are no off-shell versions \cite{Cederwall:2001bt,Cederwall:2001td,Cederwall:2001dx,Cederwall:2002df,Howe:2003cy,
Howe:2003sa,Drummond:2004vf,Drummond:2004cw}.\footnote{For a recent discussion of the non-maximal case see \cite{Cederwall:2008zv}.} The absence of auxiliary fields means that one cannot write down higher-order invariants straightforwardly; instead they have to be constructed using only the physical fields.

The construction of such invariants is the main topic of this paper. The idea is to combine the cohomological approach with the superform (``ectoplasm'') method
\cite{Gates:1997kr,Gates:1997ag,Gates:1998ec}
 of constructing invariant integrals in $d$-dimensional spacetime from closed $d$-forms in superspace. This was discussed for $d=11$ in \cite{Howe:2003cy}
and the general theory has been described in a talk \cite{Howepstalk};
some results were also given in an earlier paper
\cite{Berkovits:2006ik}.

The organisation of the paper is as follows: in the next section we briefly review the superform method, we discuss superspace cohomology in general, including cohomology groups for odd forms taking their values in $\wedge^k T_0$, where $T_0$ is the even tangent bundle, and we give some simple examples of invariants. In section 3 we discuss the cohomology groups for $N=1$ supersymmetry in $d=10$ and $11$ and show how they can be used to construct invariants. In section 4 we apply the theory to the explicit construction of various invariants in the $N=1,d=10$ heterotic superstring effective action and show that most of them, up to order $\a'^3$, can be completely determined in this manner. These $N=1,d=10$ invariants are also related
to terms in the Type I superstring effective
action by the usual heterotic-Type I duality
symmetry. We then discuss in section 5 the relation
of the invariants to pure spinor superstring scattering amplitudes, as well
as possible generalisations to
$d=11$ and $d=10,N=2$ supergravity theories.


\section{Invariant integrals}



\subsection{Ectoplasm}


In this subsection we shall briefly review the superform or ectoplasm method for constructing
superinvariants \cite{Gates:1997kr,Gates:1997ag,Gates:1998ec}. This gives an elegant and practical way of computing superspace integrals if the
dimension of spacetime is small enough and can also be useful for invariants which correspond to
subsuperspace integrals. It is closely related to the notion of rheonomy in the group manifold
approach to supersymmetry \cite{D'Auria:1982pm} and the generalized action principle \cite{sorokin:1995aa}.

Let $M$ be a supermanifold with $d$ even dimensions, and let $M_0$ denote its body. Let
$s:M_0\rightarrow M$ be a section of the projection $\p:M\rightarrow M_0$, and let $J_d$ be a
closed $d$-form on $M$. We claim that the integral

\be
 I:=\int_{M_0}\, s^* J_d\
 \la{2.1}
\ee

is independent of the choice of section, provided that we are allowed to neglect surface terms.
Since the vertical directions in $M$ correspond to the odd coordinates, it follows that the
integral $I$ is invariant under supersymmetry. Note also that $I$ is unchanged if we replace $J_d$ by
$J_d + d K_{d-1}$, so that we are really interested in the $d$th de Rham cohomology group. Note,
however, that we are only interested in forms that can be constructed from the physical fields of
the theory under discussion, so that the relevant cohomology group is $H_d^d(phys)$, and this can
be non-trivial even if the space itself is topologically trivial. Indeed, we shall assume that spacetime is topologically trivial throughout the paper.

We sketch a proof of this claim. Let $f_t$ be a one parameter family of diffeomorphisms of
$M$ generated by a vector field $v$, define the family of sections $s_t$ by $s_t=f_t\circ s$ and
set

\be
 I_t=\int_{M_0}\, s_t^* J_d= \int_{M_0}\, s^*\circ f_t^* J_d\ .
 \la{2.2}
\ee

Then

\bea
 \frac{d}{dt} I_t |_{t=0}&=&\int_{M_0}\, s^* \pounds_v J_d \w1\nn
 &=&\int_{M_0}\, s^* (d\i_v J_d + \i_v dJ_d)\nn\w1
 &=&\int_{M_0}\, d (s^* \i_v J_d)\ .
 \la{2.3}
\eea

Hence if we assume the fields in the integrand fall off fast enough at infinity, or if $M_0$ is
compact, we see that the right-hand-side vanishes thus justifying the claim. In practice, it is
usual to take $s=e$, the zero section, $e(x)=(x,0)$. In this case we have

\be
 I=\int\, d^dx\, \ve^{m_d\ldots m_1} J_{m_1\ldots m_d}(x,0)\ .
 \la{2.4}
\ee

The superspaces of interest to us have the property that their tangent bundles $T$ can be split
invariantly into even and odd parts, $T=T_0\oplus T_1$, with corresponding local bases
$E_A=(E_a,E_\a)$. If $v$ is an odd vector field, $v=v^\a E_\a$, we can identify the leading
component of $v^\a$ with the parameter of local supersymmetry, and the above discussion shows that the integral is indeed supersymmetric.

Equation \eq{2.4} allows one to
compute a component invariant integral systematically even in a curved superspace. Letting
$E_M{}^A$ denote the supervielbein which relates coordinate bases to preferred bases, and identifying the spacetime
vielbein and gravitino fields by $e_m{}^a:=E_m{}^a|$ and $\psi_m{}^\a:=E_m{}^\a |$, where the bar
denotes evaluation of a superfield at $\th=0$, we have

\bea
 I&=\int\,d^dx\,\ve^{m_1\ldots m_d}&\big( e_{m_d}{}^{a_d}\ldots e_{m_1}{}^{a_1}
 J_{a_1\ldots a_d} + d\, e_{m_d}{}^{a_d}\ldots e_{m_2}{}^{a_2} \psi_{m_1}{}^{\a_1} J_{\a_1 a_2\ldots
 a_d}+\ldots \nn\w1
 &\phantom{=\int\,d^dx\,\ve^{m_d\ldots m_1}}&
 \ldots +\psi_{m_d}{}^{\a_d}\ldots \psi_{m_1}{}^{\a_1} J_{{\a}_1\ldots {\a}_d}\big)\ ,
 \la{2.5}
\eea

where each of the $J$s is evaluated at $\th=0$.


\subsection{Cohomology of superspace}


In order to tackle the cohomology question it is useful to introduce the notion of $(p,q)$ forms,
forms with $p$ even and $q$ odd indices with respect to a preferred basis. If we let $\O_{p,q}$
denote the space of such forms\footnote{We use subscripts rather then the more usual superscripts for reasons which will become clear shortly.} we have

\be
 \O_{p,q}\ni \o=\frac{1}{p!q!} E^{\b_q}\ldots E^{\b_1} E^{a_p}\ldots E^{a_1} \o_{a_1\ldots a_p
 \b_1\ldots \b_p}\ ,
 \la{2.6}
\ee

where $E^a\, (E^\a)$ are preferred even (odd) basis forms dual to the basis vector fields $E_a\,
(E_\a)$ introduced above.

The exterior derivative splits into four parts,

\be
d=d_0 + d_1 + t_0 + t_1,
\ee

with bi-degrees
$(1,0),(0,1),(-1,2),(2,-1)$ respectively \cite{Bonora:1986ix}. It is easiest to write these using covariant derivatives
and the torsion. Thus $d_0\sim E^a(\nab_a + T_{a\cdot}{}^{\cdot})$ and $d_1\sim E^\a(\nab_\a +
T_{\a\cdot}{}^{\cdot})$ are even and odd derivatives while $t_0$ and $t_1$ are algebraic operations
involving the dimension zero and three-halves components of the torsion tensor. The operation $t_0$
applied to a $(p,q)$-form consists of contracting one of the even indices with the upper index on the
dimension zero torsion, $T_{\a\b}{}^c$, followed by symmetrisation over the $(q+2)$ odd indices.

The identity $d^2=0$ splits into various components,

\bea
 t_0^2 &=& 0\nn\w1
 d_1 t_0 + t_0 d_1 &=& 0\nn\w1
 d_1^2 + d_0 t_0 + t_0 d_0&=& 0\ ,
 \la{2.7}
\eea

together with some others which will not be needed in the following. The first of the above
equations allows us to introduce the cohomology groups $H_t^{p,q}$, the space of $t_0$-closed
$(p,q)$-forms modulo the $t_0$ exact ones \cite{Bonora:1986ix}. The groups $H_t^{0,q}:=H_t^q$ can be thought of as (generalised)
pure multi-spinors. When $N=1$ and the dimension-zero torsion takes its usual form,
$T_{\a\b}{}^{c}=-i(\c^c)_{\a\b}$, an element of this group can be represented by a
multi-spinor with $q$ symmetrised indices which is gamma-traceless on each pair. Such an object is
clearly equivalent to an $\o$ of the form

\be
 \o=\l^{\a_1}\ldots \l^{\a_q} \o_{\a_1\ldots \a_q}
 \la{2.8}
\ee

where $\l$ obeys the constraint

\be
 \l^\a (\c^a)_{\a\b} \l^\b =0\ .
 \la{2.9}
\ee

In $d=10$ such a spinor is a pure spinor in the sense of Cartan but this is not always the case.

We can also define $t_0$-cohomology groups for $(0,q)$-forms taking their values in $\wedge^k T_0$; these will turn out to be useful for finding the $H_t^{p,q}$ groups. To do this let us first define the space $\O_{p,q}^{k,l}$ consisting of $(p,q)$ forms taking their values in $\wedge^k T_0\otimes \wedge^l T_1$, i.e the space of $(p,q)$-forms which are also $(k,l)$-multivectors. The dimension-zero torsion can be made to act in two ways on this space: firstly, we define $t_0$ to act as before, i.e. ignoring the multivector indices, and secondly we define a new operation $t^0:\O_{p,q}^{k,l}\rightarrow\O_{p,q+1}^{k+1,l-1}$. In components these operations are given by

\be
 (t_0 \o)_{a_1\ldots a_{p-1},\a_1\ldots \a_{q+2}}^{b_1\ldots b_k,\b_1\ldots\b_l}=\frac{(q+1)(q+2)}{2} T_{(\a_1\a_2}{}^c \o_{|c a_1\ldots a_{p-1}|,\a_{3}\ldots \a_{q+2})}^{b_1\ldots b_k,\b_1\ldots\b_l}\ ,
 \la{2.9.1}
\ee

and

\be
 (t^0\o)_{a_1\ldots a_p,\a_1\ldots\a_{q+1}}^{b_1\ldots b_{k+1},\b_1\ldots\b_{l-1}}=
 (-1)^{p+q+1}(k+1)(q+1)\o_{a_1\ldots a_p,(\a_1\ldots\a_q}^{[b_1\ldots b_k,|\b_1\ldots\b_{l-1}\c|}T_{\a_{q+1})\c}{}^{b_{k+1}]}\ .
\ee

It is straightforward to show that $t:=t_0+t^0$ is nilpotent,

\be
 t^2=0\qquad \Leftrightarrow\qquad (t_0)^2=(t^0)^2=t_0 t^0+t^0t_0=0 \ .
 \la{2.9.2}
\ee

The operation $t$ maps $\oplus\, \O_{p-r,q+r}^{k-r,l+r}$ to $\oplus\, \O_{p-r-1,q+r+2}^{k-r,l+r}$ where the sum is over all integers $r$. We shall be interested in the cohomology groups $(H_t)^{k,0}_{0,q}:=H_t^q(\wedge^k T_0)$. Since elements of $\O_{0,q}^{k,0}$ are annihilated by $t_0$ and $t^0$, this group is given by elements of this space modulo elements of the form $t_0 \l+t^0\r$ where $\l\in \O_{1,q-2}^{k,0}$ and $\r\in\O_{0,q-1}^{k-1,1}$.

The groups $H_t^{p,q}$ will form the starting point for the analysis of the cohomology groups we
are interested in. To go further we shall define the spinorial cohomology groups $H_s^{p,q}$. To do
this we first define an odd derivative $d_s$ which acts on elements of $H_t^{p,q}$. We set

\be
 d_s [\o] :=[d_1 \o] \ ,
 \la{2.10}
\ee

where the square brackets denote equivalence classes in $H_t$ \cite{Howe:2003cy}. This definition makes sense because
$t_0$ anticommutes with $d_1$. This means that the right-hand side is unchanged if $\o\rightarrow
\o+t_0 \r$, for some $\r\in\O_{p+1,q-2}$, and also that $t_0 d_1\o=0$ because $t_0\o=0$. It is now
easy to show that $d_s^2=0$. We have

\bea
 d_s^2 [\o]&=& d_s [d_1\o]\nn\w1
 &=&[d_1^2\o]\nn\w1
 &=&[-(t_0d_0+d_0t_0)\o]\nn\w1
 &=&0\ ,
 \la{2.11}
\eea

where we have used \eq{2.7}. Given that $d_s^2=0$ we can define the cohomology groups $H_s^{p,q}$
in the obvious way: $H_s^{p,q}:= H_{d_s}(H_t^{p,q})$. The groups $H_s^{0,q}$ are isomorphic to the pure spinor cohomology groups $H_Q^q$. The latter are defined by acting on multi-pure spinors $\o$ of the type given in \eq{2.9} by $Q=\l^\a \nab_\a$. This operator squares to zero in a supergravity background if the latter obeys pure spinor integrability.

We note in passing that one can also define spinorial cohomology groups for $(0,q)$-forms taking their values in $\wedge^k T_0$. Let $h\in\O_{0,q}^{k,o}=\O_{0,q}(\wedge^k T_0)$. We can define an odd exterior derivative on such objects as follows:

\be
 (d_1 h)\cdot\o=d_1 (h\cdot \o)+(-1)^{q+1} h\cdot d_1\o\ ,
 \la{2.11.1}
\ee

where $\o\in\O_{k,0}$ and where the dot denotes contraction on all the even indices. Explicitly,

\bea
 (d_1 h)_{\a_1\ldots \a_{q+1}}^{a_1\ldots a_k}&=&(q+1) \nab_{(\a_1} h_{\a_2\ldots \a_{q+1})}+\frac{q(q+1)}{2}T_{(\a_1\a_2}{}^\c h_{|\c|\a_3\ldots\a_{q+1})}^{a_1\ldots a_k} \nn\w2
 &\phantom{=}& +\, (-1)^{q+1}k(q+1)h_{(\a_1\ldots\a_q}^{[a_1\ldots a_{k-1}|b|}T_{\a_{q+1})b}{}^{a_k]}\ .
\eea

A straightforward computation shows, provided that the dimension zero torsion is covariantly constant, that

\be
 d_1^2 h=t_0 \l + t^0 \r
 \la{2.11.2}
\ee

for some (computable) $\l\in\O_{1,q}^{k,0}$ and $\r\in\O_{0,q+1}^{k-1,1}$. We can therefore define $d_s[h]=[d_1 h]$ and $H_s^q(\wedge^k T_0)=H_{d_s}(H_t^q(\wedge^k T_0))$.

Now suppose we want to compute the de Rham cohomology group $H_d^n$; we have to find an $n$-form
$J_n$ satisfying $dJ_n=0$, modulo shifts of the form $J_n\rightarrow J_n + d K_{n-1}$. The
lowest-dimensional component of $J$ is $J_{0,n}$; it satisfies

\be
 t_0 J_{0,n}=0
 \la{2.12}
\ee

trivially and is subject to the gauge transformation

\be
 \d J_{0,n}=t_0 K_{1,n-2} \ .
 \la{2.13}
\ee

It is therefore given by an element of $H_t^{0,n}$. Now consider the $(0,n+1)$ component of $dJ=0$,

\be
 d_1 J_{0,n}+ t_0 J_{1,n-1}=0\ .
 \la{2.14}
\ee

This is equivalent to

\be
 d_s[J_{0,n}]=0\ .
 \la{2.15}
\ee

This pattern continues as the dimension is increased: the possible solutions to the de Rham
cohomology problem are generated by elements of the spinorial cohomology groups $H_s^{p,q}$ where
$p+q=n$. To find possible integral invariants using the superform method we therefore have to study
the groups $H_s^{p,q}\ ;p+q=d$. However, it should be borne in mind that not every such group
element will lead to  an integral invariant in flat superspace because it could fail to contain
$J_{d,0}$ amongst its derived components. These are the invariants we are most interested in and
will be the focus of the rest of the paper. The invariants which vanish in flat superspace will be referred to as nilpotent invariants.


\subsection{Simple examples}


A very easy example is given by $N=1,d=2$ superspace. The standard geometry is determined by
conventional constraints. The non-zero components of the torsion tensor are

\bea
 T_{\a\b}{}^c&=&-i(\c^c)_{\a\b}\nn\w1
 T_{a \b}{}^{\c}&=& (\c^c)_{\b}{}^\c S\nn\w1
 T_{ab}{}^{\c}&=&-i\ve_{ab} (\c_5)^{\c\d}\nab_\d S\ ,
 \la{2.16}
\eea

where $S$ is a scalar superfield whose components are a dimension one auxiliary field, the
gravitino field strength and the curvature scalar. The cohomology group $H_t^{0,2}$ consists of
scalar functions which can be identified with possible superspace Lagrangians. Using the freedom to
make $K$ transformations we can choose

\be
 J_{0,2}=i \c^{\natural}_{0,2} J_0\ ,
 \la{2.17}
\ee

where $\c^{\natural}_{0,2}$ denotes ``$\c_5$'' considered as a $(0,2)$-form. Similarly, we write

\be
 \c_{p,2} =\frac{1}{2. p!} E^{a_p}\ldots E^{a_1} E^{\b} E^{\a} (\c_{a_1\ldots a_p})_{\a\b}\ .
 \la{2.18}
\ee

It is easy to compute the components of $J_2$; they are

\bea
 J_{\a\b}&=& i(\c_5)_{\a\b} J_0\nn\w1
 J_{a\b}&=& (\c_5\c_a)_{\a\b} \nab^{\b} J_0\nn\w1
 J_{ab}&=& -\ve_{ab} (i\nab^{\a} \nab_{\a} J_o + 2 S J_0)\ .
 \la{2.19}
\eea

Given a particular $J_0$ we can compute its components and then find the spacetime action using
\eq{2.5}. For example, the action for a spinning string may be obtained by taking

\be
 J_0=\nab_\a X\cdot \nab^\a X\ ,
 \la{2.20}
\ee

where $X$ is the string field.

In $d=2$, for any $N$, the curvature two-form $R_{ab}$ is equal to $\ve_{ab} F$, where $dF=0$. The
supergravity action, which is purely topological as for $N=0$, is then given by setting $J_2=F$.

In $d=3$ the standard $N=1$ superspace constraints are also purely conventional, with
$T_{\a\b}{}^c=-i(\c^c)_{\a\b}$ again. In this case we need to look for closed three-forms. It is easy to see that $H_t^{0,3}=0$ and that $H_t^{1,2}$ is the space of scalar functions.
We can choose $J_{1,2}$ to be

\be
 J_{1,2}=i \c_{1,2} J_0\ ,
 \la{2.21}
\ee

where $J_0$ is the superspace Lagrangian. As in the $d=2$ case is it straightforward to compute the
component action using \eq{2.5}.

A more complicated example is provided by $d=4, N=1$ supergeometry. The groups $H_t^{0,4}$ and
$H_t^{1,3}$ are certainly not zero, but they cannot be used to generate Lagrangians forms which
give rise to non-zero invariants in flat superspace. The simplest possibility arises at the next order and makes use of a non-vanishing element of $H_t^{2,2}$ \cite{Gates:1997ag}. In two-component notation we can take

\be
 J_{ab\cdt\ddt}:=J_{\a\adt,\b\bdt,\cdt\ddt}=\ve_{\a\b}\ve_{(\adt|\cdt|} \ve_{\bdt)\ddt}J_0
 \la{2.22}
\ee

which will lead to a solution of $dJ=0$ if $J_0$ is a chiral superfield, $\bar\nab_{\adt}J_0=0$. This gives rise to a chiral invariant, i.e. an integral over chiral superspace of $J_0$. To get a full superspace integral one simply has to write $J_0$ in terms of two anti-chiral derivatives acting on some scalar superfield $S$.

The chiral example generalises straightforwardly to $d=4, N=2$ and indeed to higher $N$, although these are more complicated. There are other types of invariant in $N=2$ as can be seen from the superaction approach \cite{Howe:1981xy}
or from harmonic superspace \cite{Galperin:1984av} (see \cite{Hartwell:1994rp}
for a discussion of the relation between the two). In \cite{Biswas:2001wu}
some examples were derived from higher-rank closed forms in harmonic superspace from which one can obtain closed four-forms by integrating over the harmonic two-sphere.


\subsection{An example of a nilpotent invariant}


Consider on-shell $d=4,N=1$ supergravity in the absence of matter. The field strength superfield is a chiral dimension three-halves field $W_{\a\b\c}$ obeying

\be
 \nab_{\a} W_{\b\c\d}= C_{\a\b\c\d}\ ,
 \la{2.23}
\ee

where $C_{\a\b\c\d}$ is the totally symmetric Weyl spinor. If we set

\be
 J_{\a\b\c\d}=C_{\a\b\c\d}
 \la{2.24}
\ee

with all other components of $J$ with four spinor indices taken to vanish, then we claim that $dJ=0$ if the other non-vanishing components of $J$ are

\bea
 J_{a\b\c\d}&=&\frac{1}{4}\nab_a W_{\b\c\d}  \nn\w1
 J_{ab\c\d}&=&\ve_{\adt\bdt}( \frac{1}{4} W_{(\a\b}{}^\e W_{\c\d)\e} + \ve_{(\a|\c|}\ve_{\b)\d}W^2 ) \ ,
 \la{2.25}
\eea

where we have used the standard correspondence between a vector and a pair of spinor indices and where $W^2$ denotes the complete contraction of two $W$s. For example, the fact that $\nab_{(\a} J_{\b\c\d\e)}=0$ follows immediately from \eq{2.13}.


\subsection{Chern-Simons invariants}


The invariants that can be constructed using the superform method fall broadly into two classes: strict invariants, for which all the non-vanishing components of $J$ are tensorial, and the remainder, which can involve gauge potentials or perhaps explicit $\th$s. Many examples of invariants of the second type arise as Chern-Simons invariants, i.e. they include Chern-Simons terms together with, generically, many other tensorial terms which are required by supersymmetry. The general theory of such terms is as follows \cite{Howe:1998tsa}: let $W$ be a closed $(d+1)$-form on a superspace $M$ whose body $M_0$ has dimension $d$, and suppose that $W$ is written explicitly as $dZ$ where $Z$ is a (local) $d$-form that involves gauge potentials; if $W$ can also be written as $dK$, where $K$ is a tensorial $d$-form, then

\be
 J:=K-Z
 \la{2.26}
\ee

is a closed $d$-form which one can use to form invariants using the superform method. $Z$ obviously gives rise to the Chern-Simons or Wess-Zumino term while $K$ will give the rest of the bosonic terms which go with it. Since $W$ is a $(d+1)$-form it follows that it is exact in de Rham cohomology since the latter coincides with that of the body for a supermanifold. However, it might not be the case that it is exact for the right coefficients, which will be the space of physical fields in most examples, and therefore exactness, or Weil triviality \cite{Bonora:1986xd}, has to be checked explicitly in each case. A class of examples of this type is provided by Green-Schwarz actions for various branes \cite{sorokin:1995aa,Howe:1998tsa}; the superform method therefore provides
a rather neat explanation of the relation between the kinetic and Wess-Zumino terms in such actions. Another example is the $R^4$ invariant in $d=11$ which includes the anomaly-cancelling Chern-Simons term for the fivebrane \cite{Howe:2003cy}.


\section{$N=1$ in $d=10$ and $d=11$}


In this section we shall study the theory of the possible invariants that can arise in supergravity backgrounds in $N=1,d=10$ and $d=11$. We shall make the analysis subject to the assumption that the $H_t$ cohomology groups are determined by the allowed $p$-branes. We do not have a proof of this statement but we know of no counterexample.


\subsection{$N=1,d=10$}


The allowed $p$ branes have $p=1$ or $p=5$. These couple to two- and six-form potentials respectively and there are therefore associated closed three- and seven-form field strengths. The dimension-zero components of these are proportional to $\c_{1,2}$ and $\c_{5,2}$ respectively, both of which are annihilated by $t_0$.

The group $H_t^{0,q}:=H_t^q$ is isomorphic to the space of totally symmetric, gamma-traceless $q$-spinors, i.e. the space of pure $q$-spinors. For $p=1$ it is easy to see that $H_t^{1,0}=0$ while $H_t^{1,1}$ is equal to the space of odd vector fields. An example of this occurs in on-shell $d=10$ super Maxwell theory. If $F$ denotes the field strength two-form, $F_{0,2}=0$, so that $t_0 F_{1,1}=0$, and the solution to the latter equation being given by an element of $H_t^{1,1}$; in fact, it is the physical fermion field. There are two contributions to $H_t^{1,2}$; the first is given by $(1,2)$-forms $\o$ of the form

\be
 \o_{1,2}=\c_{1,2} f
 \la{3.1}
\ee

where $f$ is an arbitrary function. However, for $q>2$, there are no further non-trivial solutions that can be constructed using $\c_{1,2}$; this follows from the identity

\be
 (\c_a)_{(\a\b} \o_{\c_1\ldots \c_r)}=(\c^b)_{(\a\b} (\c_{ba})_{\c_1}{}^\d\o_{\c_2\ldots \c_r)\d}\ ,
 \la{3.2}
\ee

where $r=q-2$. In other words, $t_0$-closed $(1,r+2)$-forms of the type $\c_{1,2}\o_{0,r}$ are $t_0$-exact.

The remaining non-trivial possibilities are generated with the aid of $\c_{5,2}$. We find

\be
 H_t^{p,q}\supseteq H_t^{q-2}(\L^{5-p} T_0)\ ,
 \la{3.3}
\ee

provided that $p\leq 5$ and $q\geq 2$. To illustrate this consider a $t_0$-closed $(3,q+2)$-form $\o$; if it is non-trivial it can be written

\be
 \o_{3,q+2}=\c_{5,2}\l^{2,0}_{0,q}\ ,
 \la{3.3.1}
\ee

where the notation indicates that two of the even indices of $\c_{5,2}$ are contracted with the even indices on $\l$. It is easy to see that changing $\l$ by $t_0\r^{2,0}_{1,q-2}$ will lead to a $t_0$-closed change to $\o$ and it is not difficult to see that the same will be true if we change $\l$ by $t^0\r^{1,1}_{0,q-1}$. Hence the correct cohomology group is indeed $H_t^q(\wedge^k T_0)$ as claimed. If the brane assumption is correct, therefore, the non-vanishing $H_t^{p,q}$ cohomology groups for $N=1,d=10$ are

\bea
 H_t^{0,q}&=& H_t^q \nn\w1
 H_t^{1,1}&=&\O_{0,0}^{1,0}\nn\w1
 H_t^{1,q}&=&H_t^{q-2}(\L^4 T_0) + \d_{q 2}\,\O_{0,0}^{0,0},\ q\geq 2\nn\w1
 H_t^{p,q}&=&H_t^{q-2}(\L^{5-p}T_0),\ q\geq 2;\ p\in \{2,3,4,5\}\ .
\la{3.3.2}
\eea

In order to form superinvariants the possible starting points for constructing closed ten-forms are given by the cohomology groups $H_t^{p,q}$ with $p+q=10$ and $p\leq 5$. We have $H_t^{0,10}=H_t^{10}$, the space of ten-fold pure spinors, while $H_t^{p,q}\cong H_t^{q-2}(\L^{5-p} T_0)$ for $1\leq p\leq 5$. However, it turns out that only $H_t^{5,5}$ can give rise to non-nilpotent integral invariants, because it is not possible to obtain non-zero $J_{10,0}$s from any of the other possibilities. An element of $H_t^{5,5}$ can be written

\be
 J_{5,5}=\c_{5,2} f_{0,3}\ ,
 \la{3.4}
\ee

where $f_{0,3}$ determines an element of $H_t^3$. It is easy to see that the lowest-order constraint on $J_{5,5}$ coming from $dJ=0$ will be satisfied if

\be
 d_s [f_{0,3}]=0 \ ,
 \la{3.5}
\ee

or, equivalently, $f$ is in third pure spinor $Q$-cohomology group. Starting from this, one can go on to find a solution for the rest of $J$. Note that if one were to be interested in nilpotent invariants starting from some other $J_{p,q}$, for example $J_{3,7}$, then one would have to make use of the generalised spinorial cohomology groups $H_s^k(\L^l T_0)$.

In flat superspace the pure spinor cohomology groups were analysed previously in  \cite{Berkovits:2000nn}, where it was shown that the only possibilities arise for $0\leq q\leq 3$ and that $H^3_Q$ corresponds to possible actions. This provides direct confirmation of the claim made above that there are no other possibilities. On the other hand, the superform method remains valid in the presence of a non-trivial supergravity background; if we can find elements of $H_s^{0,3}$ then we can systematically find complete invariants using the superform construction.


\subsection{$N=1,d=11$}


The situation in $d=11$ is slightly simpler from the cohomological point of view since the only scalar brane is the membrane. Thus the $H_t$ groups are generated by $\c_{2,2}$ which is $t_0$ closed, but not exact, by the membrane identity. The non-trivial $H_t$ groups are

\bea
 H_t^{0,q}&\cong& H_t^q \nn\w1
 H_t^{1,q}&\cong& H_t^{q-2}(T_0),\ \ q\geq 2 \nn\w1
 H_t^{2,q}&\cong& H_t^{q-2},\ \ q\geq 2\ .
 \la{3.6}
\eea

To find a closed $11$-form we therefore need to consider $H_t^{11}$, $H_t^{8}(T_0)$ and $H_t^7$. The analysis of \cite{Berkovits:2002uc} suggests that it is the last of these groups which is important for non-nilpotent invariants. Thus we need to start with a $(2,9)$-form of the type

\be
 J_{2,9}=\c_{2,2} f_{0,7}\ .
 \la{3.7}
\ee

If $[f_{0,7}]$ satisfies

\be
 d_s [f_{0,7}]=0
 \la{3.8}
\ee

this procedure will generate a closed $11$-form and hence an invariant. Again, this analysis will be valid in the presence of a non-trivial supergravity background.


\section{Heterotic invariants}



\subsection{Conventions}


The on-shell constraints for $d=10,N=1$ supergravity were first written down in \cite{Nilsson:1981bn} and there have since been many reformulations. For our purposes
it is most convenient to describe $d=10$ supergravity in terms of the partially on-shell $128+128$ multiplet \cite{Howe:1982mt}; it is dual to the supercurrent multiplet which is conformal in the sense that the energy-momentum tensor vanishes, but which has non-local aspects \cite{Bergshoeff:1982av}. The SG multiplet consists of the graviton, the gravitino and the six-form potential $B_6$ with field strength $H_7=dB_6$; in practice we will use the dual of $H_{7,0}$ which we write as $G_{abc}$. The partial on-shell nature is reflected in the constraints which hold for the scalar curvature and the double gamma-trace of the gravitino field strength $\Psi_{ab}$. There is a fully off-shell version of the supergravity theory \cite{Howe:1982mt}, consisting of this multiplet together with two entire scalar superfields, but one of the latter has dimension $-6$ and can only be non-zero starting at order $\a'^3$.\footnote{The second scalar superfield has dimension zero and contains the dilaton and dilatino.} As we shall only be interested in the field equations up to order $\a'^2$ it is therefore reasonable to take the supergravity constraints to be those of the 128 + 128 multiplet \cite{Nilsson:1985si,Howe:1986ed}. We can study the equations of motion by introducing the two-form potential $B_2$ and its modified field strength $H_3$ satisfying $dH_3=\a'X_4:\ \ X_4:=Tr(F^2-R^2)$ where $F$ is the SYM field strength tensor; this was done for the Chapline-Manton theory in \cite{Kallosh:1985cd,Nilsson:1986md,Atick:1985de} and more generally in \cite{D'Auria:1987yv,Raciti:1989je,Bonora:1986ix,Bonora:1987xn,Bonora:1990mt,Bonora:1992tx}
(see also \cite{Bellucci:1988ff,Bellucci:1990fa,Gates:2004cd} for a different perspective, and \cite{Saulina:1995eq,Saulina:1996vn} for a dual approach).\footnote{An up-to-date summary and clarification can be found in \cite{Lechner:2008uz}.}Note, however, that at order $\a'^3$ the constraints must be changed because $H_7$ obeys the modified Bianchi identity, $dH_7=\a'^3 X_8$, where $X_8$ is an invariant eight-form constructed from $R,F$ whose form is determined by the anomaly-cancellation mechanism; indeed, it was noted that this has to be the case in \cite{Candiello:1994ew}. Moreover, it is precisely at this order that the negative-dimension auxiliary field can first be non-zero and this also implies a modification of the dimension-zero torsion \cite{Nilsson:1986cz,Howe:1986ed}.

The torsion tensor is given by

\bea
 T_{\a\b}{}^c&=&-i(\c^c)_{\a\b} \nn\w1
 T_{\a \b}{}^{\c}&=&T_{\a b}{}^{c}=0 \nn\w1
 T_{ab}{}^c&=&0 \nn\w1
 T_{a\b}{}^{\c}&=& (\c^{bc})_{\b}{}^{\c}G_{abc}+\frac{1}{6}(\c_{abcd})_{\b}{}^{\c} G^{bcd}\nn\w1
 T_{ab}{}^{\c}&=& \Psi_{ab}{}^\c\ .
 \la{4.1}
\eea

We can decompose the gravitino field-strength into irreducible, gamma-traceless components:

\be
 \Psi_{ab}=\psi_{ab} + \c_{[a} \psi_{b]} + \c_{ab}\psi\ ,
\la{4.2}
\ee

and the constraint referred to above is simply that $\psi=0$. The components of the curvature tensor are given by

\bea
 R_{\a\b,cd}&=&4i\left((\c^e)_{\a\b} G_{cde} + \frac{1}{6}(\c_{cdefg})_{\a\b} G^{efg}\right)\nn\w1
 R_{\a b,cd}&=&\frac{i}{2}(\c_b \Psi_{cd}-\c_d\psi_{bc}+\c_c\psi_{bd})_{\a}
 \la{4.3}
\eea

while the leading component of $R_{ab,cd}$ is the spacetime curvature. The curvature scalar obeys the constraint

\be
 R=12 G_{abc} G^{abc}\ .
 \la{4.4}
\ee

The non-zero components of $H_7$ are

\be
 H_{abcde\a\b}=-i(\c_{abcde})_{\a\b}\ ,
 \la{4.5}
\ee

or, more concisely, $H_{5,2}=-i\c_{5,2}$, and

\be
 H_{abcdefg}=-2\ve_{abcdefghij} G^{hij}\ .
 \la{4.6}
\ee

We shall only need the three-form to zeroth order in $\a'$. Its non-zero components are

\bea
 H_{\a\b c}&=&-iS (\c_c)_{\a\b}\nn\w1
 H_{ab \c}&=&-(\c_{ab}\chi)_{\c}\nn\w1
 \la{4.7}
\eea

as well as $H_{abc}$. The scalar field $S=\exp{2\f/3}$, where $\f$ is the dilaton, and the dilatino is defined by $\chi_\a=D_\a S$. The $H_3$ Bianchi identity also implies that

\be
 G_{abc}= \frac{1}{12} \exp(-\frac{2\f}{3}) H_{abc}
 \la{4.8}
\ee

and

\be
 D_\a \chi_\b=\frac{i}{2}(\c^a)_{\a\b}D_a S -\frac{i}{36} H_{abc}\ .
 \la{4.9}
\ee

The above equations are valid in what we shall refer to as the brane frame; the relation between the bosonic metrics in the two frames is

\be
 g_B=\exp(-\frac{2\f}{3}) g_S\ .
 \la{4.10}
\ee

At leading order the SYM field strength $F$ obeys the usual constraint that $F_{\a\b}=0$; it follows that

\be
 F_{\a b}=(\c_b \L)\a\ ,
 \la{4.11}
\ee

where $\L^\a$ is the gaugino field. We then find that

\bea
 D_\a \L^\b&=&-\frac{i}{4}(\c^{ab})_{\a}{}^{\b} F_{ab}\nn\w1
 D_\a F_{ab}&=& 2(\c_{[a}) D_{b]}\L)_\a- 2 (T_{[a} \c_{b]}\L)\a\ ,
 \la{4.12}
\eea

where $T_a$ denotes the dimension-one torsion viewed as a matrix in spin space. In principle, there could be corrections to $F_{\a\b}$ at order $\a'$, but we shall not consider these here as it is probable that such corrections vanish. At order $\a'^2$, however, there is the well-known correction which corresponds to the $F^4$ term in the Born-Infeld action \cite{Gates:1986is,Bergshoeff:1986jm}. In flat superspace it is proportional to

\be
 \stackrel{(2)}{F}_{\a\b}=\a'^2 (\c^{abcde})_{\a\b} \L\c_{abc}\L F_{de}\ .
 \la{4.13}
\ee

In a supergravity background this has to be modified slightly:

\be
 \stackrel{(2)}{F}_{\a\b}=\a'^2 (\c^{abcde})_{\a\b} \L\c_{abc}\L (e^{2\f/3}F_{de}+\frac{i}{2}\chi \c_{de} \L)\ ,
 \la{4.13.1}
\ee

in the brane frame.


\subsection{Invariants}


Higher-order corrections to the heterotic string action have been studied for many years, see, for example, \cite{Gross:1986mw,Bergshoeff:1988nn,Bergshoeff:1989de,deRoo:1992zp}. The simplest complete supersymmetric invariant that can be written down in $d=10$ is \cite{Nilsson:1986rh}

\be
 I=\int\, d^{10}x\,d^{16}\th\, E g(\f)\ ,
 \la{4.14}
\ee

where $E$ is the superdeterminant of the supervielbein. Clearly supersymmetry does not completely fix this because it is an invariant for any function $g$. It will lead to an $R^4$ term in spacetime as long as the fourth derivative of $g$ is not zero. The heterotic tree-level $R^4$ term is of this type. In the ectoplasmic approach this invariant is generated by an $f_{0,3}$ which is schematically of the form $D^{11}g(\f)$. It is interesting to note that the so-called non-minimal anomaly-free supergravity models have a non-vanishing $H_{0,3}$ which has a similar structure \cite{Lechner:1987ip,Lechner:1989kk}.

The remaining invariants we shall consider are of the Chern-Simons type and are completely fixed by supersymmetry. We shall not go into the full details here but we shall show that such invariants do indeed exist and identify the corresponding $J_{5,5}$s in some cases. There are two possible closed eleven-forms:

\be
 W^{(1)}:= \a'^3 H_3 X_8\qquad {\rm and}\qquad W^{(2)}=\a' H_7 X_4\ .
 \la{4.15}
\ee

In order to show that these define invariants of the Chern-Simons type we have to show that they can be written in the form $W^{(i)}=dK^{(i)}$ for some tensorial $K$s. Consider $W^{(1)}$: since we are only working to order $\a'^3$ we only need to know the field strengths to zeroth order. We know that $F_{0,2}=0$, but it can be shown that it is also possible to choose $R_{0,2}=0$ at this order on-shell \cite{Bonora:1986xd}. This means that the lowest non-zero component of $X_8$ will be $X_{4,4}$, and therefore the lowest component of $W^{(1)}$ is $W^{(1)}_{5,6}$. The task is to show that this can be written as $t_0 K^{(1)}_{6,4}$; if this is the case then the fact that $H_t^{p,q}=0$ for $p\geq 6$ indicates that there are no further obstructions to the existence of a suitable $K^{(1)}$.
Since $X_8$ is closed it follows that $t_0 X_{4,4}=0$ and so $X_{4,4}=\c_{5,2} X^{1,0}_{0,2}$. So

\bea
 W_{5,6}&\sim& \c_{1,2}(\c_{5,2} X^{1,0}_{0,2})\nn\w1
 &\sim& \c_{5,2}(\c_{1,2} X^{1,0}_{0,2})\nn\w1
 &\sim& t_0 (\c_{5,2} X_{1,2})\ ,
 \la{4.16}
\eea

where $X_{1,2}$ is the form obtained by lowering the upper even index on $X^{1,0}_{0,2}$ and where we have made use of $\c_{1,2}\c_{5,2}=0$. Thus the result is established; the invariant can be obtained from $J=K-Z$ by the general procedure. Since these terms include the anomaly-cancelling CS terms $Z$ it follows that they arise at one-loop in the heterotic string, but one can explicitly check that the correct factor of $e^\phi$ (i.e. no factor in the string frame) is present in the fourth-order field strength terms which arise from $K^{(1)}$.

Now let us consider $W^{(2)}$. Its lowest component is $W^{(2)}_{5,6}\sim \c_{5,2} X_{0,4}$. In this  case we know from the BPT theorem \cite{Bonora:1986ix} that $Tr(R\wedge R)$ can be written as an exact form up to a four-form whose leading component is of type $(2,2)$, from which it follows that the geometrical part of $W^{(2)}$ is guaranteed to be of the required $dK^{(2)}$ form to the order we are considering. The same will be automatically true for the $Tr(F\wedge F)$ term, up to order $\a'^2$, provided that there is no order
$\a'$ correction to $F_{0,2}$, which we assume to be the case. We therefore conclude that supersymmetric invariants can indeed be constructed from $W^{(2)}$.

In the case of $W^{(2)}$ we can consider expanding the field strengths up to order $\a'^2$, so that we can obtain invariants at first, second and third order in $\a'$. For example, at $\a'^3$ there will be terms involving $F^4$ coming from $Tr(R\wedge R)$. These can arise as follows: since $R_{0,2}\sim G$, we find that $K^{(2)}_{6,4}$ has term from $Tr(R\wedge R)$ of the form $G^2$. Now at order $\a'$ it is easy to see that there is a contribution to $G$ of the form $e^{-2\f/3}\L^2$ and hence $K^{(2)}_{6,4}$ contains $e^{-4\f/3}\L^4$. Since we need four odd derivatives to arrive at $K_{10,0}$ we will therefore obtain an $F^4$ term in the spacetime invariant. It turns out that this term is at tree level in string theory; i.e. it has  a factor of $e^{-2\f}$ in the string frame. As expected from
the analysis of
\cite{Gross:1986mw}, this term is proportional to $Tr(F^2) Tr(F^2)$
since it comes from the square of the Yang-Mills scalar $G$.
On the other hand, there is a second way of obtaining $F^4$ and that is from the $\a'^2$ deformation of $F_{0,2}$. This term is proportional
to $Tr(F^4)$ and
is actually a one-loop term and must therefore be partnered with the $B_2 X_8$ in the effective string action. This is to be expected as it is easy to see in components that the equations of motion of the two-form and six-form theories with CS terms and modified field strengths are indeed equivalent.

For both of the CS invariants we have been discussing,
the lowest component of $K$ is at least $K_{6,4}$, so that the lowest relevant component of $J$, namely $J_{5,5}=\c_{5,2}f_{0,3}$, actually comes from the CS term. We shall conclude this subsection by working out $f_{0,3}$ explicitly for the one-loop SYM $F^4$ term from both $W$s. To make life easier we shall consider an abelian gauge field and a flat supergravity background. Consider first $H_7 F^2$: the CS potential $Z$ can be chosen to be $H_7 Y_3$ where $Y$ is the CS three-form for $F^2$. Thus

\be
 J_{5,5}=\c_{5,2}Y_{0,3}\ ,
 \la{4.17}
\ee

and hence $f_{0,3}=Y_{0,3}$. Now we have to include the $\a'^2$ corrections in $F^2$, so that

\be
 Y_{\a\b\c}= A_{(\a} \stackrel{(2)} F_{\b\c)}\ ,
 \la{4.18}
\ee

and of course one takes the gamma-traceless part in the action. This expression agrees with the $f_{0,3}$ calculated explicitly from string amplitudes in the pure spinor approach as expected (see section 5).

On the other hand, in the $H_3$ version, we have $Z=H_3 X_4 Y_3$ in the abelian case with $X_4=F^2$. The lowest component of $Z$ is therefore $Z_{4,6}\sim H_{1,2} F_{1,1} F_{1,1} Y_{1,2}$. We first show that this is $t_0$ exact. We note that the only non-zero component of $H$ in flat space is $H_{1,2}\sim \c_{1,2}$ while the fact that $F_{0,2}=0$ implies that

\bea
 Y_{0,3}&=&0\nn\w1
 t_0 Y_{1,2}&=&0 \nn\w1
 d_1 Y_{1,2}+t_0 Y_{2,1}&=&0\ .
 \la{4.19}
\eea

Let us define $M:=H_3 F$; it is straightforward to show that its lowest component, $M_{2,3}$, can be written

\be
 M_{2,3}=t_0 N_{3,1}\ ,
 \la{4.20}
\ee

where $N_{abc\d}\sim (\c_{abc}\L)_\d$. Using \eq{4.19} and \eq{4.20} we find

\bea
 Z_{4,6}=-t_0(N_{3,1} F_{1,1} Y_{1,2})
 \la{4.21}
\eea

and so can be gauged away. At the next level

\be
 Z_{5,5}=H_{1,2}( (F_{1,1})^2 Y_{2,1} + 2 F_{1,1} F_{2,0} Y_{1,2})\ .
 \la{4.22}
\ee

With a little algebra it is not difficult to see that

\be
 Z_{5,5} +d_1(N_{3,1} F_{1,1} Y_{1,2})=(M_{3,2}-d_1 N_{3,1}) (F_{1,1})^2 A_{0,1} + t_0-{\rm exact}\ .
 \la{4.23}
\ee

Since $dM=0$ it follows that $M_{3,2}-d_1 N_{3,1}$ is $t_0$-closed. The exact part of this can be ignored in \eq{4.23}, and the non-trivial part is proportional to $\c_{5,2} F^{2,0}$ (where $F^{2,0}$ is $F_{2,0}$ with raised indices). The constant of proportionality cannot be zero since then $Z_{5,5}$ would also be trivial and the entire invariant would be a total derivative which it clearly is not. We also know that $F_{1,1}^2\sim \c_{5,2} L^{3,0}$, where $L^{abc}=\L \c^{abc}\L$. Hence, up to gauge terms

\be
 Z_{5,5}\sim (\c_{5,2} F^{2,0}) (\c_{5,2} L^{3,0})\ .
 \la{4.24}
\ee

If we discard all the gauge terms the resulting $Z_{5,5}$ must be $t_0$-closed, and it must be non-trivial. We can therefore choose a gauge in which

\be
 Z_{5,5}\sim \c_{5,2} f_{0,3}=\c_{5,2} A_{0,1} \stackrel{(2)} F_{0,2}\
 \la{4.25}
\ee

in agreement with the previous calculation.


\section{Invariants from superstring amplitudes}


\subsection{Heterotic superstring amplitudes}

The $N=1$ $d=10$ invariants constructed in the previous section can
easily
be verified to linearised order by comparing with heterotic
superstring scattering
amplitude computations using the pure spinor formalism.
In comparing the invariants with scattering amplitudes, it
is important to note that scattering amplitudes are computed
using vertex operators constructed from superfields which satisfy
linearised equations of motion. These linearised equations of motion
do not receive $\a'$ corrections, and are invariant under
linearised supersymmetry
transformations of the superfields.

However, the $N=1$ $d=10$
invariants of this paper are constructed using superfields
which satisfy non-linear equations of motion, and whose supersymmetry
transformations are also non-linear. This means that invariants which
are lowest order in $\a'$ and which vanish on-shell (such as the
supersymmetrisation of the Einstein-Hilbert term $\int d^{10}x \sqrt{g} R$)
are zero (or total derivatives) using the superform method with on-shell
superfields. Nevertheless, superstring scattering amplitudes will
include non-linear contributions from such invariants (such
as the tree-level scattering of three gravitons).

On the other hand, for invariants which are lowest order in $\a'$
and do not vanish on-shell (such as the
non-abelian super-Yang-Mills action), the superform method allows
the construction of the complete invariant with manifest non-linear
supersymmetry. However, scattering amplitudes will only be able
to compute contributions from these invariants
order-by-order in the linearised superfields,
and these contributions will be manifestly invariant under the linearised
supersymmetry.

In a flat background, heterotic superstring scattering amplitudes computed
using the pure spinor formalism
are expressed as

\be
{\cal A}= \langle f (\l, x,\theta) \rangle
\ee

where $f(\l,x,\theta) = \l^\a\l^\b\l^\g f_{\a\b\g}(x,\theta)$
is a superfield of ghost-number 3, and $\langle ~~\rangle$
denotes the zero mode measure factor defined such that

\be
\langle (\l\g^m\theta)(\l\g^n\theta)(\l\g^p\theta)(\theta\g_{mnp}\theta)
\rangle = 1.
\ee

When $f(\l,x,\theta)$ satisfies $Qf=0$ where $Q=\l^\a D_\a$ is
the BRST operator, the amplitude ${\cal A}= \langle f(\l,x,\theta)\rangle$
is spacetime supersymmetric. It is not difficult to verify that
the resulting invariant corresponds to $J_{5,5}=\g_{5,2} f_{0,3}$
where $f_{0,3} = f_{\a\b\g}$. So the superform method for constructing
invariants is directly related to the zero-mode measure factor in
the pure spinor formalism.

For three-point heterotic massless tree amplitudes,

\be
{\cal A}= (\a')^{-2}\langle V_1(z_1) V_2(z_2) V_3(z_3) \rangle
\ee

where
$V = \bar c \l^\a [A_{\a I}(x,\theta) J^I + B_{\a m}(x,\theta)\bar\partial x^m]$
is the massless vertex operator, $\bar c$ is the right-moving
reparameterization ghost, $J^I$ are the right-moving currents
for the gauge group, and $A_{\a I}$ and $B_{\a m}$ are
the linearised superfields for super-Yang-Mills and supergravity.
When $A_{\a I}$ and $B_{\a m}$ satisfy the equations of motion

\be
(\g^{mnpqr})^{\a\b} D_\a A_{\b I} =
(\g^{mnpqr})^{\a\b} D_\a B_{\b m} = 0, \quad \partial^m B_{\b m}=
\partial^n\partial_n A_{\b I} =
\partial^n\partial_n B_{\b m} = 0,
\ee

the vertex operator $V$ is BRST-closed with respect to the left
and right-moving BRST operators, $Q=\int dz~\l^\a d_\a$ and
$\bar Q = \int d\bar z(\bar c \bar T + b\bar c \bar \partial c)$
where $\bar T$ is the right-moving stress tensor.
It is easy to check that these equations for $A_{\a I}$ and
$B_{\a m}$ describe the linearised on-shell super-Yang-Mills
and supergravity fields.

After integration over the right-moving worldsheet fields, one obtains
the expression
${\cal A}= \langle f(\l,x,\theta) \rangle$ where

\bea
f_{\a\b\g} &=& \a' Tr (A_\a^1 A_\b^2 A_\g^3) +
\a' (k^2 \cdot B_\a^1)(k^3 \cdot
B_\b^2)(k^1 \cdot B_\g^3) \nn\w2
 &\phantom{=}& + \a' (B_\a^1 \cdot k^2) Tr(A^2_\b A^3_\g) +
\a' (B_\a^2 \cdot k^3) Tr(A^3_\b A^1_\g) +
\a' (B_\a^3 \cdot k^1) Tr(A^3_\b A^1_\g) \nn\w2
 &\phantom{=}&  +
(B_\a^1 \cdot B_\b^2)(k^1\cdot B^3_\g) +
(B_\a^2 \cdot B_\b^3)(k^2\cdot B^1_\g) +
(B_\a^3 \cdot B_\b^1)(k^3\cdot B^2_\g)\ .
\eea

The terms in $f_{\a\b\g}$ proportional to $\a'$ correspond to the cubic
on-shell terms in $W^{(2)}$ of equation (4.15), whereas the terms
independent of $\a'$ come from the cubic terms in the Einstein-Hilbert
action. For example, $\a' Tr(A_\a^1 A_\b^2 A_\g^3)$ is the onshell
contribution to the Yang-Mills Chern-Simons term $Y_{0,3}$
in equation \eq{4.17}.

Since three-point massless amplitudes do not receive $\a'$
corrections,
one needs to consider
higher-point amplitudes to check invariants which are higher orders in $\a'$.
For example, the four-point one-loop amplitude has
been shown to lowest order in $\a'$ to be proportional to \cite{berk}
${\cal A} = \langle \l^\a\l^\b \l^\g f_{\a\b\g}(x,\theta)\rangle $
where

\be
f_{\a\b\g} = Tr[A_\a^1 (\g^{mnpqr})_{\b\g} (\Lambda^2 \g_{mnp}\Lambda^3)
F_{qr}^4] + permutations.
\ee

Comparing with \eq{4.13}, one sees that $J_{5,5} =\g_{5,2} f_{0,3}$
where $f_{0,3} = Tr (A_\a F^{(2)}_{\b\g})$ as desired.


\subsection{$N=2$ $d=10$ invariants}


Just as $N=1$ $d=10$ invariants are related to scattering amplitudes
for heterotic or open superstrings, $N=2$ $d=10$ invariants are
related to scattering amplitudes of Type II superstrings. Since vertex
operators of Type II superstrings can be expressed as left-right
products of open superstring vertex operators, it is natural to propose
that $N=2$ $d=10$ invariants are related to superforms

\be
f_{0,3,3} \sim f_{\a\b\g\ah\bh\gh}
\la{5.7}
\ee

where
the first subscript denotes the number of vector indices, the
second subscript denotes the number of unhatted ``left-moving''
spinor indices, and
the third subscript denotes the number of hatted ``right-moving''
spinor indices.
For the Type IIA superstring, hatted and unhatted spinor indices
have opposite chirality, while for the Type IIB superstring, they
have the same chirality.

Type II superstring amplitudes using the pure spinor formalism
are expressed as

\be
{\cal A} = \langle f \rangle
\ee

where
$f=\l^\a\l^\b \l^\g \lh^{\ah} \lh^{\bh} \lh^{\gh} f_{\a\b\g\ah\bh\gh}$
is BRST closed with respect to $Q=\int dz \l^\a d_\a$ and
$\bar Q = \int d\bar z \hat\l^{\ah} \hat d_{\ah}$,
$\l^\a$ and $\lh^{\ah}$ are pure spinors satisfying
$\l\g^a\l = \lh\g^\a\lh=0$, and

\be
\langle (\l\g^m\theta)(\l\g^n\theta)(\l\g^p\theta)(\theta\g_{mnp}\theta)
(\lh\g^q\hat\theta)(\lh\g^r\hat\theta)
(\lh\g^s\hat\theta)(\hat\theta\g_{qrs}\hat\theta)
\rangle = 1.
\la{5.9}
\ee

So, at least to linearised level, there is a supersymmetric $N=2$ $d=10$
invariant associated with $f_{0,3,3}.$
For the $N=2A$ case, one can relate this invariant
with the superform method by defining

\be
J_{0,5,5} = \g^{abcde}_{(\d\e} f_{\a\b\g)(\ah\bh\gh}
(\g_{abcde})_{\hat\delta \hat\e)},
\la{5.10}
\ee

which is the component of lowest dimension of
a closed 10-form as we shall show in subsection (5.4). However,
an analogous construction does not work for the $N=2B$ case
since $\g^{abcde}_{\d\kappa} (\g_{abcde})_{\hat\d\hat\kappa}=0$ if
the hatted and unhatted spinor indices have the same chirality.

An alternative method for constructing $N=2A$ $d=10$ invariants
is to
dimensionally reduce $N=1$ $d=11$ invariants. Using the pure spinor
version of the $d=11$ supermembrane, amplitudes are computed as
${\cal A} = \langle f \rangle$ where
$f$ has ghost-number seven and the zero model measure factor
is of the form $\langle (\theta)^9 (\l)^7 \rangle=1$.
The resulting $d=11$ invariant is obtained from the superform
whose component of lowest dimension is
$J_{2,9} = \g_{2,2} f_{0,7}.$ This invariant can be obtained directly as a CS invariant starting from the closed twelve-form $W=G_4 X_8$, where $G_4$ is the supergravity four-form field strength and $X_8$ the anomaly-cancelling $R^4$ eight-form \cite{Howe:2003cy}. It is reasonable to assume that the lowest non-vanishing component of $K$ is $K_{3,8}$ , so that $J_{2,9}=-Z_{2,9}$. From this we conclude that it is possible to take $f_{0,7}$ to be proportional to $Y_{0,7}$ where $dY_7=X_8$.

To dimensionally reduce, suppose that
the $d=11$ superform $J_{n+1,10-n}= J_{a_1 ... a_{n+1} ~\a_1 ... \a_{10-n}}$
is independent of the $d=11$
coordinate $x^{11}$. Then it is easy to show that

\be
J^{d=10}_{n,10-n} \equiv J^{d=11}_{11 ~a_1 ... a_{n} ~\a_1 ... \a_{10-n}}
\ee

is a closed 10-form which defines an $N=2A$ $d=10$
invariant. Note that $J^{d=11}_{n,11-n}$ vanishes for $n<2$, so
$J^{d=10}_{n,10-n}$ vanishes for $n<1$ when defined using this method.

An obvious question is how to relate these superstring
and supermembrane constructions of $N=2A$ $d=10$ invariants.
The relation is not obvious since the supermembrane method constructs
an invariant whose component of lowest dimension is $J_{1,9}$,
whereas the invariant
constructed using the superstring method has $J_{0,10}$ as its
component of lowest dimension. Furthermore, $J_{0,10}$ constructed
using the superstring method has five hatted and five unhatted indices,
whereas the supermembrane-derived $J_{1,9}$ naively has no restriction
on the relative number of hatted and unhatted spinor indices.


\subsection{Proposal for $N=2$ invariants using doubled superspace}

In the previous subsection, we presented two approaches for
constructing $N=2A$ invariants using the superform method.
However, despite the fact that Type IIA and Type IIB superstrings
are related by T-duality, neither of these approaches appear to be useful
for constructing $N=2B$ invariants. In this subsection, we present
an alternative proposal for constructing $N=2$ invariants which works
equally well for $N=2A$ and $N=2B$. However, we have not yet
checked the consistency of this proposal at the non-linear level.

This alternative proposal for constructing $N=2$ supersymmetric invariants
involves doubling both the number of bosonic
coordinates and fermionic coordinates; it was first discussed in \cite{Isaev:1988qn,Isaev:1988hg}. The bosonic coordinates
in this ``doubled'' superspace
will be called $x^m$ and $\hat x^{\hat m}$ for $m,\hat m=0$ to 9,
and the fermionic
coordinates will be called $\theta^\mu$ and $\hat\theta^{\hat\mu}$
for $\mu,\hat\mu=1$ to 16.
One then defines the supervierbeins $E_M{}^A$
and $E_{\hat M}{}^{\hat A}$ where $M=(m,\mu)$ and $\hat M = (\hat m, \hat\mu)$
describe the curved coordinates and $A=(a,\a)$ and $\hat A=(\hat a,\hat\a)$
describe the tangent-space indices.
In a flat background, the only non-vanishing torsions will be
defined to be
$T_{\a\b}{}^a = -i(\g^a)_{\a\b}$ and
$T_{\ah\bh}{}^{\hat a} =-i (\g^{\hat a})_{\ah\bh}$.

Such a doubled formalism has previously appeared in discussions of T-duality
for the heterotic superstring \cite{siegel,siegelhet} as well as in
the Type II superstring. Although we have not yet checked
the consistency of the latter construction at the non-linear level (see \cite{Isaev:1990eq} for some partical results on this),
we think it is a promising possibility for constructing $N=2$
$d=10$ invariants using the superform method.

Superforms $f_{A_1 ... A_{10}; \hat A_1 ... \hat A_{10}}$ in this ``doubled''
superspace contain 10 unhatted indices and
10 hatted indices, where $A$ is either $a$ or $\a$ and
$\hat A$ is either $\hat a$ or $\hat\a$.
Furthermore, all components $f$ of this superform are constrained to satisfy

\be
{\del\over {\del x^m}} f = {\del \over {\del \hat x^{\hat m}}} f
\ee

so that $f$ only depends on 10 bosonic and 32 fermionic coordinates.
It is easy to see that if $f$ is a closed 20-form satisfying the
above constraint the integral

\be
\int d^{10} x|_{\hat x=x}~
\epsilon^{m_1 ... m_{10}} ~\epsilon^{\hat m_1 ... \hat m_{10}}
~f_{m_1 ... m_{10}; \hat m_1 ... \hat m_{10}}
\ee

is invariant under $N=2$ $d=10$ supersymmetry where
$\int d^{10} x_{\hat x=x}$ means that one is to set $\hat x^{\hat m}=x^m$ and
integrate over $x^m$. Note that this invariant can also
be written as

\be
\int d^{10} x ~\int d^{10} \hat x ~
\epsilon^{m_1 ... m_{10}} ~\epsilon^{\hat m_1 ... \hat m_{10}}
~f_{m_1 ... m_{10}; \hat m_1 ... \hat m_{10}}
\ee

where $x^m -\hat x^{\hat m}$ is defined
to take values on a compact space of unit volume.

Using this doubled superspace, we conjecture that
a closed $(10;10)$ superform can be constructed such that
the component of $f$ with lowest dimension is

\be
f_{5,5;5,5} = \g_{5,2;0,0} ~\g_{0,0;5,2}~ f_{0,3;0,3}
\ee

where $\g_{5,2;0,0} = \g^{m_1 ... m_5}_{\a\b}$,
$\g_{0,0;5,2} = \g^{\hat m_1 ... \hat m_5}_{\ah\bh}$, and $f_{0,3;0,3}$
is defined in Eq. \eq{5.7}. At the linearised level, it is easy
to see that such a definition would lead to the invariants
described by Type II superstring amplitudes using the measure factor
of Eq. \eq{5.9}. For example, just as the cubic super-Yang-Mills
amplitude is described by $f_{0,3}= A_\a A_\b A_\g$ where $A_\a$
is the spinor super-Yang-Mills gauge potential, the
cubic contribution to $N=2$ supergravity is described by
$f_{0,3;0,3} = B_{\a\ah} B_{\b\bh} B_{\g\gh}$ where $B_{\a\ah}$
is the spinor-spinor component of the antisymmetric tensor superfield
which can be interpreted as the ``left-right'' product of two
spinor super-Yang-Mills gauge potentials.


\subsection{Remarks on $N=2,\ d=10$ cohomology}


Provided that the $H_t^{p,q}$ cohomology groups are determined by scalar branes, for $N=2,d=10$ it follows that these will only be non-zero for $p=0,1$, since there are only strings in this case. Consider a $t_0$-closed $(1,q)$-form $\o$ in IIA; it can be written

\be
 \o_{1,q}=\C^{\natural}_{1,2} \m_{0,q-2}\ ,
 \la{6.1}
\ee

where $\C^{\natural}_{1,2}$ denotes $\C_{a}\C_{11}$ viewed as a $(1,2)$-form and where $\C$ denotes the $32\xz 32$ Dirac matrices. Now $\o$ will change by a $t_0$ exact term if $\m$ does, but it will also do so if $\m$ is changed by $\C^{\natural}_{0,2}\r_{0,q-4}$. This means that the $d=10$ IIA group $H_t^{1,q}$ is isomorphic to the $d=11$  group $H_t^{0,q-2},\ q\geq2$. This explains how dimensional reduction from $d=11$ to $N=2,d=10$ works cohomologically because the relevant groups are still the $d=11$ ones.

In IIB there are two strings and hence two possible $(1,2)$-forms which can be used to write down elements of $H_t^{1,q}$.

There are other groups we can consider in $N=2$ as there is now a triple-grading of forms: $\O_{p,q}=\sum_{r+s=q}\O_{p,r,s}$, where the $(r,s)$ labels correspond to unhatted and hatted odd indices. We also have

\bea
 t_0&=&\t_0 + \hat\t_0 \w1\nn
 d_1&=&\del_1+ \hat\del_1\w1\nn
 t_1&=&\t_1 +\hat\t_1\ ,
 \la{6.2}
\eea

where the tri-degrees are $(-1,2,0)$ ($(-1,0,2)$ for $\t_0$ ($\hat\t_0$), $(0,1,0)$ ($(0,0,1)$ for $\del_1$ ($\hat\del_1$) and $(2,-1,0)$ ($(2,0,-1)$) for $\t_1$ ($\hat\t_1$). In principle one could also have a component of $t_0$ with tri-degree $(-1,1,1)$ but it vanishes in on-shell supergravity. One can easily write out $d^2=0$ in terms of these operations. There are various cohomology groups that can be constructed. For example, one can define $H_\t^{p,r,s}$, the space of $\t_0$-closed $(p,r,s)$-forms modulo the exact ones. Since $\del_1\t_0+\t_0\del_1=0$ and $\del_1^2 +\t_0 d_0+d_0\t_0=0$ we can define spinorial cohomology groups of the form $H_s^{p,r,s}$, and similarly for the hatted sector. However, in general it does not seem to be very easy to analyse the spinorial cohomology groups we are interested in terms of these partial ones.

We shall give one example, in IIA, which relates to the $J_{0,5,5}$ discussed in
Eq.\eq{5.10}. Suppose we have a $(0,r,s)$-form $\o_{0,r,s}$, with $r,s\geq3$, and suppose that

\be
 \o_{0,r,s}=\C_{0,2,2} f_{0,r-2,s-2}\ ,
 \la{6.3}
\ee

where

\be
 \C_{0,2,2}:=\c_{5,2,0} \c^5{}_{,0,2}\ ,
 \la{6.4}
\ee

i.e. the contraction of two five-index gamma-matrices considered as a $(0,2,2)$-form. We would like to construct a closed $(r+s)$-form $\o$ starting from $\o_{0,r,s}$. We can solve the first non-trivial component of $d\o=0$ if

\bea
 \del_1\o_{0,r,s}+\t_0 \o_{1,r-1,s}&=&0\w1\nn
 \hat\t_0 \o_{1,r-1,s}&=&0\ ,
 \la{6.5}
\eea

and similarly for the hatted components. The first of these is satisfied if

\be
 \del_1 f_{0,r-2,s-2}+\t_0 f_{1,r-3,s-2}=0\ .
 \la{6.6}
\ee

If this is so, then

\bea
 \del_1\o_{0,r,s}&=&-\C_{0,2,2}\t_0 f_{1,r-3,s-2} \w1\nn
 &=&-(\c_{5,2,0}\c^5{}_{,0,2})(\c_{1,2,0} f^1{}_{,r-3,s-2})\ .
 \la{6.7}
\eea

Now we can use the fact that $\c_{5,2,0}\c_{1,2,0}=0$ to shuffle the indices so that the even index on $\c_{1,2,0}$ is contracted with one of the even indices of $\c^5{}_{,0,2}$. This means that we can choose $\o_{1,r-1,s}$ to be

\be
 \o_{1,r-1,s}=\c_{5,0,2} g^4{}_{,r-1,s-2}\ ,
 \la{6.8}
\ee

where

\be
 g_{4,r-1,s-2} \sim \c_{5,2,0} f^1{}_{,r-3,s-2}\ .
 \la{6.9}
\ee

It is now immediate that $\o_{1,r-1,s}$ defined by \eq{6.8} is annihilated by $\hat\t_0$. Similar considerations apply to $\o_{1,r,s-1}$.



\section{Discussion}


In this paper, we combined the superform method with pure spinor
cohomology to construct invariants with manifest $N=1$ $d=10$
supersymmetry. This method was used to construct on-shell invariants
of the heterotic superstring effective action up to order $(\a')^3$
corrections including supersymmetrisation of the $F^2$, $R^2$, $F^4$, $R^4$,
$BF^4$ and $BR^4$ terms.
Although we did not attempt to expand these invariants in terms
of component fields it should be straightforward to perform
this component expansion by evaluating the $J_{n,10-n}$ forms
which appear in the invariants.

There are several possible generalisations of our results. One
obvious one is to construct invariants for terms in
the effective action which are higher-order in $\a'$. It would
be very interesting to identify restrictions imposed by supersymmetry
which constrain the possible couplings to the dilaton. Since the dilaton
counts loops, these restrictions might be used for testing string
duality conjectures as in \cite{sethi:1999su}.

Another possible generalisation is to use the superform method
to construct $d=11$ and $N=2$ $d=10$ invariants. As discussed in
section (5.2), there are some unresolved puzzles concerning the
relation of these invariants, and it would also be of interest
to investigate further the doubled superspace proposal of section (5.3).

\vskip 20pt

{\bf Acknowledgements:} We would like to thank
Carlos Mafra for useful discussions
and the Isaac Newton Institute
for Mathematical Sciences
for their hospitality during the workshop on Strong Fields,
Integrability and Strings.
NB would like to thank
CNPq grant 305814/2006-0 and
FAPESP grant 04/11426-0
for partial financial support. This work was also supported in part by EU grant (superstring theory) MRTN-2004-512194.

\end{document}

    [ Part 3, Application/PDF  276KB. ]

    [ Unable to print this part. ]

    [ Part 3, Application/PDF  309KB. ]

    [ Unable to print this part. ]